\documentclass[epj-spec]{svjour}
\usepackage{graphics}

\newcommand{\be}{\begin{equation}}
\newcommand{\ee}{\end{equation}}
\newcommand{\ba}{\begin{eqnarray}}
\newcommand{\ea}{\end{eqnarray}}
\newcommand{\ban}{\begin{eqnarray*}}
\newcommand{\ean}{\end{eqnarray*}}

\newcommand{\sandwich}[3]{\mbox{$ \langle #1 | #2 | #3 \rangle $}}
\newcommand{\ket}[1]{\mbox{$ | #1 \rangle $}}
\newcommand{\bra}[1]{\mbox{$ \langle #1 | $}}

\newcommand{\si}{\sigma}
\newcommand{\demi}{\frac{1}{2}}
\newcommand{\compl}{\begin{picture}(8,8)\put(0,0){C}\put(3,0.3){\line(0,1){7}}\end{picture}}

\newcommand{\one}{\leavevmode\hbox{\small1\normalsize\kern-.33em1}}

\newcommand{\Tr}{\mbox{Tr}}

\begin{document}
\title{Entanglement and Irreversibility in the Approach to Thermal Equilibrium}
\subtitle{Known and New Results on Thermalizing Quantum Channels
for Qubits}
\author{Valerio Scarani\inst{1}\fnmsep\thanks{\email{physv@nus.edu.sg}}}
\institute{Quantum Information Science and Technology, Physics
Department, National University of Singapore, 2 Science Drive 3, Singapore 117542}
\abstract{When a physical system is put in contact with a very large thermal bath, it undergoes a dissipative (i.e., an apparently irreversible) process that leads to thermal equilibrium. This dynamical process can be described fully within quantum physics, involving only unitary, therefore reversible, maps. The information, initially present in the system, is not erased, but is diluted in the bath because of entanglement. Irreversibility may arise if, after quantum information has been thus diluted, some classical information is lost. This paper reviews a model for thermalization that displays these features. Two new analytical results are provided for the zero-temperature channels: a new quantitative measure of entanglement, and a study of irreversibility in the case where the lost classical information is the label of the particles in the bath.} 
\maketitle
\section{Introduction}
\label{intro}

A well-established tenet of statistical physics says that an
ensemble of physical systems in thermal equilibrium with a large
reservoir (``canonical ensemble'') will show a statistical
behavior: a state of energy $E$ will be occupied with a
probability proportional to $e^{-\beta E}$ where
$\beta=\frac{1}{k_BT}$, $T$ is the temperature and $k_B$ is Boltzmann's constant.
Statistical physics, whose fruitfulness is beyond question, takes
the existence of statistical ensembles as a starting point. The
question about the origin, and even the meaning, of these
statistics, is ultimately still unsettled. Recent remarkable
developments have shown that ensembles are somehow
generic in a {\em kinematic} sense: for instance, if one picks at random a state $\ket{\Psi}$ in the
Hilbert space describing a large number $N$ of particles, then, under suitable constraints,
the state of much smaller sub-systems
$\rho_n=\Tr_{N-n}\ket{\Psi}\bra{\Psi}$ shall almost always be close to a
canonical state \cite{japan,mahler,mahlerbook,popescu,goldstein}.

Here, we consider rather a \textit{dynamical} process, the
approach to thermal equilibrium, or \textit{thermalization}, a
typical example of dissipative phenomenon\footnote{Some terminological issues to avoid confusion: I use \textit{dissipation} as the phenomenological fact that energy, or information, which was initially concentrated in some physical system, flows through the evolution into the systems that interact with the first one. When a great number of degrees of freedom is involved, dissipation leads to apparent (and practical) \textit{irreversibility}; but contemporary physics does not contemplate any fundamentally irreversible process.}. It is defined as
follows. A large number $N$ of particles are already in the
canonical state $\xi$; one brings along a new particle, prepared
in an arbitrary state $\rho$, possibly pure: thermalization is the process, at the end of which the new particle reaches arbitrarily close to the canonical state, while the bath is almost unmodified. The naive description of this process, \ba \rho\otimes \xi^{\otimes
N}&\longrightarrow & \xi^{\otimes (N+1)}\,,
\label{nonunitary}\ea is clearly non-unitary, because input orthogonal
states are mapped on the same final state. For sure, non-unitary maps are allowed in quantum physics as descriptions of open systems: specifically, (\ref{nonunitary}) is possible if the bath itself is coupled to an environment, to which it transfers all the information about $\rho$. Remarkably, quantum physics allows to find \textit{unitary} maps that involve only the bath and the new particle, and that describe thermalization as well. This possibility is due to \textit{entanglement}: the information, which was initially concentrated in $\rho$, is not lost in an external environment, but is encoded differently, being spread between the system and the bath in a coherent
and reversible way. As a result, one may have \ba \rho\otimes
\xi^{\otimes N}\longrightarrow \sigma&\mbox{ such that
}&\Tr_N\sigma\approx \xi\,. \label{unitary}\ea This means that the
final state is not the exact thermal state of $N+1$ particles
$\xi^{\otimes (N+1)}$; however, when one picks any particle out,
its partial state is as close as desired to the thermal state. In
other words, $\Tr(\hat{A}\sigma)$ will correspond to the expected
thermal average for all single-particle observables
$\hat{A}=\frac{1}{N+1}\sum_{k=1}^{N+1}A_k$, with $A_k$ the
operator acting as $A$ on particle $k$ and trivially on the
others. In short: entanglement allows to construct a dissipative channel with fully reversible dynamics. The study of such channels is richer, and probably more satisfactory from the standpoint of physics, than the simple acceptance of the naive map (\ref{nonunitary}).

In this paper, I review the thermalizing channels that were
presented a few years ago for a specific model of the system and
the bath \cite{sca02}. This approach to thermalization, that can
be seen in the broader context of ``quantum homogenization''
\cite{zim02,zim05}, was proposed as a benchmark for exploring a
quantitative link between entanglement and dissipation. I also
present two new results for this model: a new measure of
entanglement, and an analytical computation for a numerical result
obtained in \cite{zim02} for a model of irreversibility.

\section{The Model and the Results}
\label{sec:1}

\subsection{Definition of the Model}
\label{ssmodel}

The model is defined by the following assumptions on the
kinematics and dynamics.

\textit{Kinematics and Free Dynamics.} The particle to be
thermalized (the ``system'' $S$ hereafter) is a two-level system, i.e. a qubit. The thermal
bath is a reservoir composed of an arbitrary large number $N$ of
qubits, which interact with an external field but not with one another (a model
featuring such interactions has been studied in Ref.~\cite{dawson}). The free Hamiltonian for the bath is therefore $H_B= \sum_{i=1}^{N}h[i]$ where
$h[k]$ is the operator acting as $h=-E\si_z$ on the qubit $k$ and
trivially on the other qubits. We denote the projectors on the
eigenstates of $\si_z$ by $P_0=\ket{0}\bra{0}$ and
$P_1=\ket{1}\bra{1}$. The initial state of the system is
arbitrary. The single-particle equilibrium state is \ba \xi&=&
e^{-\beta h}/\mbox{Tr}\big(e^{-\beta h}\big)\,=\, pP_{0}+qP_{1}
\label{statebath} \ea with $p=\demi(1+\tanh (\beta E))$ and
$q=1-p$. We set $E>0$, so that $\ket{0}$ is the ground state, and
$p=1$ corresponds to $T=0$. The initial state of the bath is the
thermal state $\rho_B\,=\, e^{-\beta H_B}/\mbox{Tr}\big(e^{-\beta
H_B}\big)=(\xi)^{\otimes N}$.

\begin{figure}
\begin{center}
\resizebox{0.4\columnwidth}{!}{\includegraphics{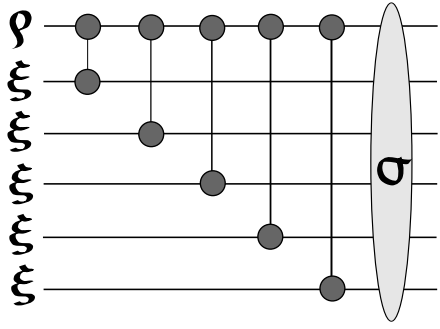} }
\caption{The collisional model for thermalization: quantum circuit representation. The input (left) shows
system qubit in an arbitrary state $\rho$, bath qubits in the thermal state $\xi$. The unitary operations connect sequentially the system qubit with each of the qubits in the bath. At the output (right), the qubits are in a correlated state $\sigma$. Thermalization is achieved if $\Tr_N\sigma\approx \xi$ holds for all the single-qubit states, whatever the initial state $\rho$ of the system qubit.}
\label{fig:model}       
\end{center}
\end{figure}

\textit{Interaction System-Bath.} We consider a \textit{collision
model}, in which the system interacts sequentially with the qubits in the bath one by one. Each step of this stroboscopic evolution is described by a hamiltonian $H$, or the
corresponding evolution $U=e^{iH}$, acting on
$\compl^2\otimes\compl^2$. Moreover, we consider that a qubit of the bath undergoes
at most one interaction with the system (Fig.~\ref{fig:model}), so
the input state of the ancilla is always $\xi$. The evolution of
the system is finally described by the iteration of the
completely-positive (CP) map $T_{\xi}$ defined as \ba
\rho^{(k+1)}&=&\mbox{Tr}_B\big[ U\,(\rho^{(k)}\otimes
\xi)\,U^{\dagger}\big]\,\equiv\,T_{\xi}\,[\rho^{(k)}]\,.
\label{rhoout} \ea

\subsection{Requirements for a Thermalizing Channel}

We want to find all the two-qubit unitary operations that define a
thermalizing channel in our scenario. Four requirements are
suggested by physics:

\textit{First Requirement.} $U$ should not depend on temperature (i.e. on
$p$), but may depend on the label $z$ of the local free
Hamiltonian. The requirement is motivated by the fact that $z$ is
a locally available information, while temperature is a
characteristic of an ensemble. Also, we want to characterize
interactions that thermalize at all temperatures, avoiding
possible pathological examples which would define the correct
physics only for a specific value of this parameter. So from now
on we write $U_z$.

\textit{Second Requirement.} If the system is prepared in the
local equilibrium state $\xi$, nothing should happen, because the
thermalization is already achieved. Formally: \ba
U_z\,(\xi\otimes\xi)\,U_z^{\dagger} = \xi\otimes\xi &\;\;\mbox{
for }&\xi = p P_0 + q P_1\,.\label{req1}\ea Note that this is stronger than requiring $\xi$ to be a fixed point of
$T_{\xi}$, because we ask that the qubit of the bath is unchanged too. Because of the first requirement, (\ref{req1}) should hold for all $p$.

\textit{Third Requirement.} For any input state $\rho$ of the
system, the iteration of the map $T_{\xi}$ leads to
thermalization:\ba \rho^{(n)}=T_{\xi}^n[\rho]\,\longrightarrow
\xi\;\;\forall\,\rho\,. \label{req2} \ea

\textit{Fourth Requirement.} For any input state $\rho$ of the
system, the fluctuations introduced in the state of the bath are
small.

\subsection{Solution: All Thermalizing Channels for the Model}

The combination of the first and the second requirements imply
that the subspaces $P_0\otimes P_0$, $P_1\otimes P_1$ and
$P_0\otimes P_1+ P_1\otimes P_0$ must be invariant under the
action of $U_z$. To
prove this assertion, we notice that on the l.h.s. of (\ref{req1})
the term $U_z\,P_0\otimes P_0\,U_z^{\dagger}$ appears with the
weight $p^2$, the term $U_z\,\big(P_0\otimes P_1+ P_1\otimes
P_0\big)\,U_z^{\dagger}$ with the weight $p(1-p)$, and the term
$U_z\,P_1\otimes P_1\,U_z^{\dagger}$ with the weight $(1-p)^2$.
Since we want condition (\ref{req1}) to hold for all $p$, the
three subspaces must be separately invariant. Thus the first and
the second requirement restrict $U$ to take the form \ba
\begin{array}{lcl}
\ket{00} & \longrightarrow & e^{i\chi_{0}}\ket{00}\\
\ket{11} & \longrightarrow & e^{i\chi_{1}}\ket{11}\\
\ket{01} & \longrightarrow & e^{i\chi_{2}}\left[\cos\phi\ket{01}+e^{i\varphi_2}\sin\phi\ket{10}\right]\\
\ket{10} & \longrightarrow &
e^{i\chi_{3}}\left[\cos\phi\ket{10}-e^{-i\varphi_2}\sin\phi\ket{01}\right]\,.
\end{array}\label{generalu}
\ea Remarkably, each of these unitaries defines a thermalizing
channel (in other words, the third requirement is automatically
fulfilled when enforcing the first and the second). To prove this
assertion, we compute explicitly $\rho^{(n)}=T_{\xi}^n[\rho]$ as a
function of the parameters of the initial state $\rho$. Let's
write \ba \rho^{(n)}&=&d^{(n)}\, P_0\,+\,(1-d^{(n)})\,
P_1\,+\,k^{(n)}\ket{0}\bra{1}\,+\,k^{(n)*}\ket{1}\bra{0}\,. \label{statein} \ea
Inserting the explicit form (\ref{generalu}) for $U_z$ into
(\ref{rhoout}), we find that the effect of the map $T_\xi$ is
given by $d^{(n+1)}=d^{(n)}\cos^2\phi+p\sin^2\phi$ and
$k^{(n+1)}=\cos\phi\,\lambda \,k^{(n)}$ with
$\lambda=pe^{i(\chi_0-\chi_3)} + qe^{i(\chi_2-\chi_1)}$ (note that
$|\lambda|\leq 1$). A straightforward iteration gives $d^{(n)}$
and $k^{(n)}$ as a function of the parameters $d^{(0)}$ and
$k^{(0)}$ of the initial state $\rho$: \ba
d^{(n)}&=&[1-(\cos\phi)^{2n}]\,p\,+\,(\cos\phi)^{2n}\,d^{(0)}\,,
\label{itert}\\ k^{(n)}&=& k^{(0)}\,\big(\lambda\,\cos\phi\big)^n
\,. \label{iterk}\ea Thus, whenever $\phi\neq 0$, the iteration of
$T_{\xi}$ yields $d^{(n)}\rightarrow p$ and $k^{(n)}\rightarrow
0$, i.e. thermalization (\ref{req2}). Finally, the fourth
requirement is fulfilled by taking $\phi$ small enough. In fact,
it can be verified that the fidelity
of the state $\sigma$ of a bath qubit after the interaction with
respect to the thermal state satisfies
$F=\Tr\Big[\big(\xi^\demi\sigma\xi^\demi\big)^\demi\Big]\geq
\cos\phi$ for all temperatures and for all input states of the system.

Of the six free parameters defining the most general $U_z$
(\ref{generalu}), only $\phi$ and the differences $\chi_0-\chi_3$
and $\chi_2-\chi_1$ define the CP-map $T_{\xi}$. The reason is
that there are two symmetries of the physical process of
thermalization. The first one is the usual choice of a global
phase; the second is the freedom of choosing the global phases of
$\ket{0}$ and $\ket{1}$ for qubits in the bath, both before and
after the interaction with the system. Mathematically, if $U_z$
defines a CP-map $T_{\xi}$, then, writing $u(x)=P_0+e^{ix}P_1$ a rotation along the $z$
axis, $ e^{i\chi}\big[\one\otimes
u(\alpha)\big]\,U_z\, \big[\one\otimes u(\beta)\big]$ defines the
same CP-map for all choices of $\alpha$ and $\beta$. So much for the three parameters which play strictly no
role. In addition, the phase of $\lambda$ is a measurable
parameter but does not add any insight on the thermalization
process. In fact, this parameter is associated to a rotation
around the $z$ axis, or in other words, to a redefinition of the
$x$ and $y$ axes of the Bloch sphere in the plane perpendicular to
$z$, for the system qubit. Since there is nothing, in the physics
of the model, that would single out a specific axis in the $(x,y)$
plane, we can study thermalization for any specific choice of the
phase of $\lambda$.

All in all, the physics of thermalization in this model can be
studied on a two-parameter family of unitary transformations. A
convenient choice is\footnote{Here the sign of $\theta$ is the
opposite than in Ref.~\cite{sca02}.} \ba V_z(\phi,\theta)&=&
\begin{array}{lcl}
\ket{00} & \longrightarrow & e^{i\theta}\ket{00}\\
\ket{11} & \longrightarrow & e^{i\theta}\ket{11}\\
\ket{01} & \longrightarrow & \cos\phi\ket{01}+i\sin\phi\ket{10}\\
\ket{10} & \longrightarrow & \cos\phi\ket{10}+i\sin\phi\ket{01}
\end{array}\label{specificu}
\ea which can be written as $e^{i\theta/2}e^{iH(\phi,\theta)}$
with \ba
H(\phi,\theta)&=&\demi\,\left[\phi\left(\sigma_x\otimes\sigma_x+
\sigma_y\otimes\sigma_y\right)+\theta
\sigma_z\otimes\sigma_z\right]\,. \ea

\subsection{The Physics of Thermalizing Channels}

In spite of the extreme simplicity of the model, the thermalizing
channels display a rich physics. Here is a review of these
features. For the known results, the demonstration is given in
Refs \cite{sca02,zim02}; the new results will be demonstrated in
Section \ref{sec:2} below.

\textit{The partial swap.} The particular choice $\theta=\phi$
makes $V_z(\phi,\phi)\equiv V(\phi)$ independent of $z$: this is
the only interaction, in this model, that is independent not only
of temperature, but also on the parameter of the local
Hamiltonian. This unitary can be written as \ba V(\phi)&=&
\cos\phi\one+i\sin\phi U_{swap} \ea where $U_{swap}=V(\pi/2)$ is
the operation that swaps the state of the two qubits:
$\ket{\psi_1}\ket{\psi_2}\rightarrow \ket{\psi_2}\ket{\psi_1}$.
The mechanism for thermalization is quite intuitive in this case.
With a channel made of pure swaps, the system would be in state
$\xi$ after one single interaction, but this would introduce a
large fluctuation in the bath. The partial swap, in the meaningful
limit $\cos\phi\approx 1$, is the gentle version of this process:
after many steps, the state of the system has been ``swapped''
into the bath, but the fluctuations in the bath itself are small.

\textit{Dissipation and Decoherence.} The dynamics of
thermalization is described in (\ref{itert}) and (\ref{iterk}),
with $\lambda=pe^{i\theta} + qe^{-i\theta}$. The parameter
$\theta$ appears only in the dynamics of the off-diagonal term,
hence is entirely related to decoherence. The re-equilibration of
populations, i.e. dissipation, is governed by $\phi$
alone\footnote{Note that $\phi$ has to appear in both the diagonal
and off-diagonal term, as is indeed the case: any re-equilibration
of populations must be balanced by a sufficient amount of
variation in the coherence, in order to avoid evolution into
non-physical states (matrices with negative eigenvalues).}.
Interestingly, dissipation and decoherence are decoupled in this
model, in the sense that
$V_z(\phi,\theta)=V_z(\phi,0)V_z(0,\theta) =
V_z(0,\theta)V_z(\phi,0)$. Also, one can verify that \ba
V_z(\phi,\theta)&=& V_z(0,\theta-\phi)V(\phi)\,:\ea all the
thermalizing unitaries in this model can be seen as the
application of the partial swap followed by an additional
reduction of coherence.

\textit{Continuous-time limit.} In order to pass from the discrete
dynamics indexed by $n$ to a continuous-time dynamics with
parameter $t$, one sets $n=t/\tau_0$, and lets the interaction
time $\tau_0$ go to zero together with $\phi$ and $\theta$,
keeping constant the dissipation rate
$\frac{\phi^2}{\tau_0}=\frac{1}{T_1}$ and the phase fluctuation
rate $\frac{2\theta^2}{\tau_0}=\frac{1}{T_{pf}}$. One finds that
in the continuous-time limit, the processes of dissipation
(\ref{itert}) and decoherence (\ref{iterk}) are exactly
exponential: \ba
d(t)&=&e^{-t/T_1}d(0)+(1-e^{-t/T_1})p\,,\\
|k|(t)&=&e^{-t/T_2}|k|(0) \ea with
$\frac{1}{T_2}=\frac{1}{2\,T_1}+p\,q\,\frac{1}{T_{pf}}$. For
$\theta=0$ or at zero temperature, the bound $T_1\geq \demi T_2$
(see e.g. \cite{alicki}, p. 120) is saturated. Note also that it is possible to relate this dynamics to a master equation of the Lindblad type \cite{zim05}.

\textit{Dissipation and Entanglement.} As we mentioned in the
introduction, this model was proposed as a benchmark for studying
the link between entanglement and dissipation. This link has been
explored along several directions. An especially strong link is
provided by the following observation: we have seen that any
thermalizing unitary is equivalent to one of the $V(\phi,\theta)$
up to local unitaries (LU), and it can be shown that all the
$V(\phi,\theta)$ are inequivalent under LU \cite{sca02,kraus}.
Thus $\phi$ and $\theta$ (or any equivalent choice of two
parameters) are necessary and sufficient to define all the
properties of entanglement; but in turn, these parameters uniquely
define the relaxation times $T_1$ and $T_2$. So, for the model
under study, the relaxation times are directly related to
entanglement.

In Refs \cite{sca02,zim02,zim03} several kinds of entanglement
have been computed: the largest amount of two-qubit entanglement
that a single application of $V(\phi,\theta)$ can generate, the
entanglement of a given qubit versus all the others etc. These are
measures of bipartite entanglement, the only ones which were
available at the moment of writing. Computable measures of
\textit{multipartite} entanglement have been proposed since
\cite{buch1,buch2,yusong}. According to one of these measures,
after interaction of the system with $n$ qubits in the bath, an
amount of entanglement\footnote{For comparison, the same measure
of entanglement applied to the $(n+1)$-qubit GHZ state gives
${\cal E}_{n+1}(GHZ)=\sqrt{2}\sqrt{1-2^{-n}}$.} \ba {\cal
E}_{n+1}&=& 2|c_1|\,\sqrt{(1-c^{2n})
\left(1-|c_1|^2\,\frac{1-c^{2(n+1)}}{1+c^{2}}\right)}\,,\label{new1}\ea
with $c=\cos\phi$, is generated at $T=0$ for an arbitrary pure
initial state $c_0\ket{0}+c_1\ket{1}$ (subsection \ref{ss21}).
This expression is monotonically increasing in $n$; for any given
$n$, the maximal amount of entanglement is generated by the input
state $\ket{1}$, which is indeed the farthest from equilibrium.
For this case $c_1=1$, ${\cal
E}_{\infty}=2\frac{c}{\sqrt{1+c^{2}}}\rightarrow \sqrt{2}$ in the
limit of weak interactions.

\textit{Irreversibility.} The thermalization process that we
described is certainly reversible, since it is described by the
unitary operation ${\cal U}=U_{S,n}...U_{S,2}U_{S,1}$. To study
irreversibility, one must add some lack of knowledge. In our
context, the most natural way of doing this is to suppose that the
labels of the qubits in the bath are randomly permuted before
reversing the evolution ${\cal U}$, and we don't control this
permutation. Then the reversed evolution reads ${\cal
U}_{\pi}={\cal U}^{\dagger}\Pi{\cal U}$ which is equal to the
identity if and only if $\Pi=\one$. Note that $\Pi$ is unitary,
and consequently so is ${\cal U}_{\pi}$: apparent irreversibility is due to the fact
that we reverse only ${\cal U}$, because we are not supposed to
control $\Pi$. In Ref.~\cite{zim02}, we have provided numerical
results for this effect. In subsection \ref{ss22} below, we compute analytically the average fidelity $\bar{F}$ of the reconstructed state with respect to input state of the system qubit, for the case $T=0$ and for the specific evolution $V(\phi,\theta=0)$. Taking the excited state $\ket{1}$ as input, one has
\ba
\bar{F}&\approx & \frac{1}{n}+\frac{4[c/(1-c)]^2}{n(n-1)}
\label{new2}\ea
with $c=\cos\phi$, for $n$ large enough; the expression for an arbitrary input state is Eq.~(\ref{avfid}). From these analytical results, one learns that the expected decrease of the average fidelity is much slower than the decrease in $1/n!$ of the probability of retrieving the initial state exactly.

\section{Demonstration of the New Results}
\label{sec:2}

This is a technical section, devoted to the demonstration of the
two new results. The common starting
point is the form of the $(n+1)$-qubit state obtained after
interaction of the system qubit with $n$ qubits in the bath. We
consider an arbitrary pure input state
$\ket{\psi}=c_0\ket{0}+c_1\ket{1}$; at $T=0$, the equilibrium
state is $\xi=P_0$ pure as well, so the initial state is
$\ket{\Psi_{in}}=\ket{\psi}_S\ket{0...0}_B$ and the state stays
pure under the evolution. After the system has interacted with $n$
qubits in the bath, the state reads \ba {\cal
U}_n\ket{\Psi_{in}}&=&
c_0e^{in\theta}\ket{0}_S\ket{0^n}_B\,+\,c_1\Big[ c^n
\ket{1}_S\ket{0^n}_B + is\ket{0}_S\Big(\sum_{k=1}^n
c^{k-1}e^{i(n-k)\theta}\ket{1_k}_B\Big)\Big] \label{state}\ea
where $\ket{0^n}_B$ and $\ket{1_k}_B$ are $n$-qubit product states
of all $\ket{0}$, respectively of $\ket{1}$ for qubit $k$ and
$\ket{0}$ for the others. Through all this section, we use
$c=\cos\phi$, $s=\sin\phi$ and $B=\{1,...,n\}$.

\subsection{Amount of multipartite entanglement}
\label{ss21}

We want to compute the multipartite entanglement in the
$(n+1)$-qubit pure state (\ref{state}). The measure of
multipartite entanglement that we are going to compute is Eq.~(6)
of \cite{buch1}, Eq.~(88) of \cite{buch2}: \ba {\cal E}_{n+1} &=&
2^{1-\frac{n+1}{2}}\,\sqrt{2^{n+1}-2-S_{n+1}} \label{En}\ea where
is $S_{n+1}$ is the sum of $\Tr(\rho_i^2)$ over all partial
traces. Since the state is pure, the spectral properties of
$\rho_{S,b}$ and of $\rho_{B\setminus b}$ are identical for all
subset $b$ of $B$; therefore we must compute $S_{n+1}= 2
\sum_{b\neq B} \Tr\rho_{S,b}^2$. Now, from (\ref{state}) one has
that all the $\rho_{S,b}$ are rank two states of the form
$\ket{\varphi_{S,b}}\bra{\varphi_{S,b}} +
x_{b}\ket{0...0}\bra{0...0}$ with $\ket{\varphi_{S,b}}$ a
non-normalized state whose precise form is not important here, and
with \ba x_{b}=1-||\varphi_{S,b}||^2= |c_1|^2s^2\sum_{k\not\in b}
c^{2(k-1)}\,.\ea So the quantity to be computed is \ba S_{n+1}&=&
2 \sum_{b\neq B}\left[(1-x_b)^2+x_b^2\right]\,=\, 2^{n+1}-2 -
4\big[\sum_{b\neq B}x_b -\sum_{b\neq B} x_b^2\big]\,.\label{Sn}\ea
To compute the first sum, notice that
$\frac{1}{|c_1|^2s^2}\sum_{b\neq B}x_b= \sum_{b\neq B}
\sum_{k\not\in b} c^{2(k-1)}\,=\, \sum_{k=1}^n c^{2(k-1)}N_k$
where $N_k$ is the number of sets $b$ such that $k\not\in b$.
Clearly $N_k=2^{n-1}$ for all $k$; therefore we are left with a
geometrical series and finally \ba \sum_{b\neq B}x_b&=&
2^{n-1}\,|c_1|^2\,\left(1-c^{2n}\right)\,. \label{somme1}\ea The
second sum that we have to compute for (\ref{Sn}) is \ba
\sum_{b\neq B} x_b^2&=& |c_1|^4s^4\,\Big[ \sum_{b\neq B}
\sum_{k\not\in b} c^{4(k-1)} + \sum_{b\neq B} \sum_{k,k'\not\in
b|k\neq k'} c^{2(k+k'-2)}\Big]\,\equiv\,|c_1|^4s^4(I_1+I_2)\,. \ea
Proceeding exactly as above, one obtains
$I_1=2^{n-1}\frac{1-c^{4n}}{s^2(1+c^2)}$. Similarly, \ba I_2&=&
\sum_{k=1}^{n}\sum_{k'\neq
k}c^{2(k+k'-2)}N_{kk'}\,=\,2^{n-1}\,\frac{c^2(1-c^{2(n-1)})(1-c^{2n})}{s^4(1+c^2)}
\ea because $N_{kk'}=2^{n-2}$ for all $k$ and $k'\neq k$, and we
have applied to $x\equiv c^2$ the generic formula\footnote{This
formula can be re-derived from the usual sum of geometric series by noticing that
$\sum_{k=0}^{n-1}\sum_{k'\neq k} x^{k+k'}$ is equal to
$2\sum_{k=0}^{n-2}\sum_{k'=k+1}^n x^{k+k'}$, or alternatively to
$\big(\sum_{k}x^k\big)^2-\sum_{k}x^{2k}$.} \ba
\sum_{k=0}^{n-1}\sum_{k'\neq k} x^{k+k'}&=&
\frac{2x(1-x^{n-1})(1-x^{n})}{(1-x)^2(1+x)}\,.\label{generic}\ea
Inserting everything back into (\ref{Sn}), then into (\ref{En}),
one obtains (\ref{new1}).

\subsection{Irreversibility via random permutations in the reservoir}
\label{ss22}

We start again from state (\ref{state}). After the interaction
with $n$ qubits of the bath, we keep the system qubit and apply a
random permutation $\pi:B\longrightarrow B$ to the indices of the
qubits in the bath. At this point, the state reads \ba \Pi{\cal
U}_n\ket{\Psi_{in}}&=&
c_0e^{in\theta}\ket{0}_S\ket{0^n}_B\,+\,c_1\Big[ c^n
\ket{1}_S\ket{0^n}_B + is\ket{0}_S\Big(\sum_{k=1}^n
c^{\tilde{\pi}(k)-1}e^{i(n-\tilde{\pi}(k))\theta}\ket{1_k}_B\Big)\Big]
\ea with $\tilde{\pi}=\pi^{-1}$. On this state we apply the
reverse evolution, at the end of which we find \ba {\cal
U}_n^{\dagger}\Pi{\cal U}_n\ket{\Psi_{in}}&=&
\big(c_0\ket{0}+c_1f_{\pi}\ket{1}\big)_S\ket{0^n}_B\,+\,
c_1\sqrt{1-|f_{\pi}|^2}\,\ket{0}_S\sum_k\alpha_k\ket{1_k}_B\ea
where we don't need the explicit form of the $\alpha_k$, but we
need \ba f_{\pi}&=&c^{2n}+s^2\sum_{k=1}^n
c^{k+\tilde{\pi}(k)-2}\,e^{i\theta[k-1-\tilde{\pi}(k-1)]}\,.
\label{eff}\ea Let $\rho^{(n_{\rightarrow},\pi,n_{\leftarrow})}=\Tr_B\left[{\cal
U}_{\pi}\ket{\psi,0...0}\bra{\psi,0...0}{\cal
U}_{\pi}^{\dagger}\right]$ be the reconstructed state of the system qubit: its fidelity with respect to the initial state is \ba F(\pi)\,\equiv\,
\sandwich{\psi}{\rho_S^{(n_{\rightarrow},\pi,n_{\leftarrow})}}{\psi}
&=& |c_0|^2+
|c_1|^2\,\left[|f_{\pi}|^2+2|c_0|^2(\mbox{Re}f_{\pi}-|f_{\pi}|^2)\right]\,.\label{fidelity}\ea To quantify
irreversibility, we must now average over all possible
permutations. At this stage, we simplify the problem by focusing
only on the unitary operation $V(\phi,\theta=0)$, so that
$f_{\pi}$ becomes real\footnote{This is not the only
simplification: if $\theta\neq 0$, expression (\ref{eff}) contains
both $\tilde{\pi}(k-1)$ and $\tilde{\pi}(k)$.}. To find
$\bar{F}=\frac{1}{n!}\sum_{\pi}F(\pi)$ we have to compute \ba
\bar{f_{\pi}}&=& c^{2n}+s^2\frac{1}{n!}\sum_{\pi}\sum_{k=1}^n
c^{k+\tilde{\pi}(k)-2}\,\equiv\, c^{2n}+s^2 I_1\,,\\
\bar{f_{\pi}^2}&=&c^{4n}+2c^{2n}s^2 I_1 +
s^4\frac{1}{n!}\sum_{\pi}\Big(\sum_{k=1}^n
c^{k+\tilde{\pi}(k)-2}\Big)^2 \,\equiv\, c^{4n}+2c^{2n}s^2 I_1 +
s^4I_2\,.\ea The pattern for these calculations is quite similar
to the one we used in subsection \ref{ss21}. Consider first $I_1$:
observing that there are $(n-1)!$ permutations such that
$\tilde{\pi}(k)=j$, one can decouple the two sums and write \ba
I_1&=& \frac{(n-1)!}{n!}\Big(\sum_{k=1}^n c^{k-1}\Big)
\Big(\sum_{j=1}^n
c^{j-1}\Big)\,=\,\frac{1}{n}\left(\frac{1-c^n}{1-c}\right)^2\,.\ea
Similarly, using in particular (\ref{generic}), we obtain \ba
I_2&=&\frac{1}{n!}\sum_{\pi}\sum_{k=1}^n c^{2(k+\tilde{\pi}(k)-2)}
+ \frac{1}{n!}\sum_{\pi} \sum_{k,k'|k\neq
k'}c^{k+k'+\tilde{\pi}(k)+\tilde{\pi}(k')-4}\nonumber\\
&=& \frac{1}{n}\left(\frac{1-c^{2n}}{1-c^2}\right)^2+
\frac{1}{n(n-1)}\left(\frac{2c(1-c^{n-1})(1-c^{n})}{(1-c)^2(1+c)}\right)^2\,.
\ea So finally the average fidelity for the reconstructed state is
\ba \bar{F}&= & |c_0|^2+
|c_1|^2\,\left[\bar{f_{\pi}^2}+2|c_0|^2(\bar{f_{\pi}}-\bar{f_{\pi}^2})\right]\, \approx\, |c_0|^2+
|c_1|^2\,\left[s^4I_2+2|c_0|^2(s^2I_1-s^4I_2)\right]\,.\label{avfid}\ea Inserting the expressions above, one finds that the convergence $\bar{F}\longrightarrow |c_0|^2$ is indeed present as expected, but is very slow (remember that $c=\cos\phi$ is close to 1 in the meaningful limit). The expression (\ref{new2}) given above is obtained by setting $c_0=0$ and $c^n\approx 0$.

\section{Conclusions and Open Questions}

In conclusion, we have reviewed a collisional model for the dynamical process of thermalization (or more generally, homogenization). In spite of its simplicity, the model exhibits a rich variety of interesting physical features, e.g. decoupling of dissipation and decoherence and exponential decays in time for both ($T_1$ and $T_2$). Moreover, this kind of model provides a benchmark to study quantitative links between the parameters of thermalization and the entanglement generated by the evolution. This relation has been studied in this paper using a new measure of multipartite entanglement. Also as expected, if classical information (here, the labels of the particles in the bath) is lost after the entanglement has been distributed, the channel becomes irreversible. Here, this irreversibility has been quantified analytically for an example of channel.

Here is a list of open questions related to thermalization:

\begin{itemize}

\item Staying within the model, the two new results presented here
are partial, and suggest themselves lines for further study. First of all, both results have been derived only for the case $T=0$, because in this case the state of all the qubits is pure. In addition, the
measure of multiparticle entanglement (\ref{En}) has no
clear-cut operational meaning\footnote{It is known that this and similar
measures of multipartite entanglement may detect entanglement
beyond the bipartite case \cite{yusong}. We note also that, contrary to
what is stated in \cite{dawson}, the result obtained in
\cite{zim03} does not exclude the existence of multipartite
entanglement in our thermalizing channels.} and, as it turned out,
does not grasp all the physics of entanglement in the model, because the
resulting quantity (\ref{new1}) is independent of $\theta$.
Similarly, the irreversibility has been studied only for the
channel with $\theta=0$, and in the case where only the bath
qubits are permuted but the system qubit is known.

\item The studied model has some remarkable features: e.g., the
fact that the third requirement is automatically implied by the
two others; the fact that (up to local unitaries) thermalizing
channels are characterized by two parameters in one-to-one
correspondence with the relaxation times $T_1$ and $T_2$; etc. Are
these generic features of collisional thermalizing channels, or
just artefacts of the choice of dealing with qubits?

\item We discussed in the text the intuitive character of the
partial swap. This operation is only possible because the system
is of the same dimensionality as the particles in the bath. What
happens if one wants to thermalize a $d$-dimensional quantum
system with a bath of (say) qubits? Can the qubits be used one by
one, or must one take them in bunches?

\end{itemize}

\begin{acknowledgement}

I am very grateful to the organizers of the 382. Wilhelm und Else Heraeus Seminar (Bad Honnef, 8-10.01.2007) for the invitation to attend this very stimulating workshop and give a talk there. This article benefits from discussions with Andreas Buchleitner, Daniel Burgarth, Oscar Dahlsten, Dai Li and Peter Zoller.

\end{acknowledgement}


\begin{thebibliography}{}

\bibitem{japan} H. Tasaki, Phys. Rev. Lett. \textbf{80}, (1998) 1373

\bibitem{mahler} J. Gemmer, A. Otte, G. Mahler, Phys. Rev. Lett. \textbf{86}, (2001) 1927

\bibitem{mahlerbook} J. Gemmer, M. Michel, G. Mahler, \textit{Quantum Thermodynamics -
Emergence of Thermodynamic Behavior within Composite Quantum
Systems}, LNP 657 (Springer Verlag, Berlin, New York, 2004)

\bibitem{popescu} S. Popescu, A.J. Short, A. Winter,  Nature Physics  \textbf{2}, (2006) 754

\bibitem{goldstein} S. Goldstein, J.L. Lebowitz, R. Tumulka, N. Zangh\`{i},
Phys. Rev. Lett. \textbf{96}, (2006) 050403

\bibitem{sca02} V. Scarani, M. Ziman,
P. \v{S}telmachovi\v{c}, N. Gisin, V. Bu\v{z}ek, Phys. Rev. Lett.
\textbf{88}, (2002) 097905

\bibitem{zim02} M. Ziman,
P. \v{S}telmachovi\v{c}, V. Bu\v{z}ek, M. Hillery, V. Scarani, N.
Gisin, Phys. Rev. A \textbf{65}, (2002) 042105

\bibitem{zim05} M. Ziman,
P. \v{S}telmachovi\v{c}, V. Bu\v{z}ek, Open systems and
information dynamics \textbf{12}, (2005) 81

\bibitem{dawson} C.M. Dawson, A.P. Hines, R.H. McKenzie, G.J.
Milburn, Phys. Rev A \textbf{71}, (2005) 052321

\bibitem{alicki} R. Alicki, K. Lendi, \textit{Quantum Dynamical Semigroups and Applications},
Lecture Notes in Physics (Springer Verlag, Berlin, 1987)

\bibitem{kraus}
B. Kraus, J.I. Cirac, Phys. Rev. A \textbf{63}, (2001) 062309

\bibitem{zim03} M. Ziman,
P. \v{S}telmachovi\v{c}, V. Bu\v{z}ek, J. Opt. B: Quantum
Semiclass. Opt. \textbf{5}, (2003) S439

\bibitem{buch1} F. Mintert, M. Ku\'s, A. Buchleitner, Phys. Rev.
Lett. \textbf{95}, (2005) 260502

\bibitem{buch2} F. Mintert, A.R.R. Carvalho, M. Ku\'s, A. Buchleitner, Phys.
Rep. \textbf{415}, (2005) 207 

\bibitem{yusong} C. Yu, H. Song, Phys. Rev. A \textbf{73}, (2006) 022325


\end{thebibliography}
\end{document}